\documentclass[vci-paper]{elsart5p}
\usepackage{amssymb}
\usepackage{epsfig}

\graphicspath{{ViennaPaper}}

\newcommand{\m}{\mathrm{m}}
\newcommand{\mm}{\mathrm{mm}}
\newcommand{\cm}{\mathrm{cm}}

\begin{document}

\begin{frontmatter}

\title{Scintillator Tile Hadron Calorimeter with Novel SiPM Readout }

\author{M.~Danilov 
(representing the CALICE collaboration)} 
\address{Institute of Theoretical and Experimental Physics,
117218 Moscow, Russia}

\begin{abstract}
The CALICE collaboration is presently constructing a test hadron
calorimeter (HCAL) with 7620 scintillator tiles read out by novel
photo-detectors - Silicon Photomultipliers (SiPMs).  This prototype is
the first device which uses SiPMs on a large scale.  We present the
design of the HCAL and report on measured properties of more than 10
thousand SiPMs.
We discuss the SiPM efficiency, gain, cross-talk, and noise rate
dependence on bias voltage and temperature, including the spread in
these parameters.  We analyze the reasons for SiPM rejection and
present the results of the long term stability studies. The first
measurements of the SiPM radiation hardness are presented.  We compare
properties of SiPM with the properties of similar devices, MRS APD and
MPPC.  A possibility to make the tiles thinner and to read them out
without WLS fibers has been studied.

\end{abstract}

\end{frontmatter}

\section{Introduction}
{\label{introduction} 

The physics requirements at the International Linear Collider (ILC)
impose high demands on the performance of calorimeters. The ultimate
goal is to achieve a jet energy resolution of about $30\%/\sqrt{E}$,
in order to increase the sensitivity for reconstruction of the W, Z
and Higgs bosons and supersymmetric particles.  Monte Carlo studies
have indicated that this goal can be achieved by utilizing the
Particle Flow (PF) method.  In this approach energies of neutral
particles (photons, neutrons, and $K_L$) are measured in calorimeters
while charged tracks are measured with a better precision in the
tracker.  The showers produced by charged particles should be removed
from the calorimetric measurements. This method requires a very high
calorimeter granularity in order to reconstruct showers produced by
neutral particles in a vicinity of showers produced by charged
particles. The PF method defines to a large extent the whole
architecture of the ILC detector.  So far the PF method is studied
mainly with MC. These studies demonstrated that a very high
granularity of about $3\times3~\cm^2$ is required. Such a granularity
can be achieved with the novel photo-detectors developed in Russia,
Silicon Photomultipliers (SiPMs) \cite{sipm}.  The CALICE
collaboration is presently constructing a test hadron calorimeter
(HCAL) with 7620 scintillator tiles read out by SiPMs.  This
calorimeter, together with a silicon tungsten electromagnetic
calorimeter (ECAL) and a scintillator strip Tail Catcher and Muon
Tracker (TCMT) will test the PF detector concept with hadron and
electron beam data. It is the first large scale application of the
SiPMs. Operational experience will be extremely important for the
design of a few hundred times larger ILC calorimeter.  New types of
Multipixel Geiger Photo-Diodes (MGPD) are being developed now by
several firms including Hamamatsu.  Their comparison is important for
the development of the best photo-detectors for the ILC HCAL.

\section{The Hadron Calorimeter}
{\label{prototype}

The HCAL is a 38-layer sampling calorimeter made of a
plastic-scintillator steel sandwich structure with a lateral dimension
of about $\rm 1 \times 1\m^2$.  Each layer consists of 1.6~cm thick
steel absorber plates and a plane of 0.5~cm thick plastic scintillator
tiles housed in a steel cassette with two 2~mm thick walls.  The tile
sizes vary from $3 \times 3 \rm ~\cm^2$ for $10 \times 10$ tiles in
the center of the module, to $6 \times 6 \rm ~\cm^2$ in the
intermediate region ($4 \times 24$ tiles), and $12 \times 12 \rm
~\cm^2$ ($4 \times 5$ tiles) in the outer region.  In the last eight
layers, the granularity is decreased to $6 \times 6 \rm ~\cm^2$ in the
central region to save the number of channels.  Each tile has a
$1~\mm$ diameter wavelength-shifting (WLS) fiber inserted into a 2~mm
deep groove. The fiber is coupled to a SiPM via an air gap. To
increase the light yield, the other fiber end is covered with a
mirror. The grooves have a quarter-circle shape in the smallest tiles
and a full-circle shape in the other tiles. The sides of each tile are
matted to provide a diffuse reflection. The tile faces are covered
with a 3M Superradiant foil.

 The readout electronics was developed by the Orsay and DESY groups
\cite{delataille}.  A VME-based data acquisition system was produced
by the UK Calice group \cite{dauncy}.  The HCAL is equipped with a LED
based calibration system produced in Prague~\cite{Vrba}. It provides
light signals up to 200~Minimum Ionizing Particle (MIP) equivalent
individually to each tile.

\section{Silicon Photomultipliers}
{\label{sipm}

A SiPM is a multipixel silicon photodiode operated in the Geiger mode
\cite{sipm}. These detectors were developed and manufactured by
MEPhI/PULSAR in Russia. The SiPM photosensitive area is $\rm 1.1\times
1.1~\mm^2$. It holds 1156 pixels, with the size of $\rm 32\times \rm
32~\mu m^2$. SiPMs are reversely biased with a voltage of $ \sim 50$~V
and have the gain $ \sim 10^6$. Once a pixel is fired it produces the
Geiger discharge. The analog information is obtained by summing up the
number of fired pixels. So the dynamic range is limited by the total
number of pixels. Each pixel has a quenching resistor of the order of
a few $\rm M\Omega$ built in, which is necessary to break off the
Geiger discharge.  Photons from a Geiger discharge in one pixel can
fire neighboring pixels. This leads to a cross talk between pixels.
The pixel recovery time is of the order of 100~ns.  With smaller
resistor values the recovery time reaches about 20~ns and a pixel can
fire twice during the pulse from the tile. This leads to a dependence
of the SiPM saturation curve on the signal shape which is not
convenient.  The SiPMs are insensitive to magnetic fields which was
tested up to 4 Tesla \cite{sipm}.

More than 10000 SiPMs have been produced by the MEPhI/PULSAR group and
have been tested at ITEP. The tests are performed in an automatic
setup, where 15 SiPMs are simultaneously illuminated with calibrated
light from a bundle of Kuraray Y11 WLS fibers excited by a UV LED
. During the first 48 hours, the SiPMs are operated at an increased
bias voltage, that is about 2~V above the normal operation
voltage. Next, the gain, noise and relative efficiency with respect to
a reference photomultiplier are measured as a function of the bias
voltage. The bias voltage working point is chosen as the one that
yields 15 pixels for a MIP-like signal provided by the calibrated LED.

At the working point, we measure several SiPM characteristics. With
low-light intensities of the LED, we record pulse height spectra that
are used for the gain calibration.  A typical pulse height spectrum is
shown in figure~\ref{fig:sipm-pixel}, in which up to 9 individual
peaks corresponding to different number of fired pixels are clearly
visible.  
\begin{figure}
\includegraphics[width=35mm]{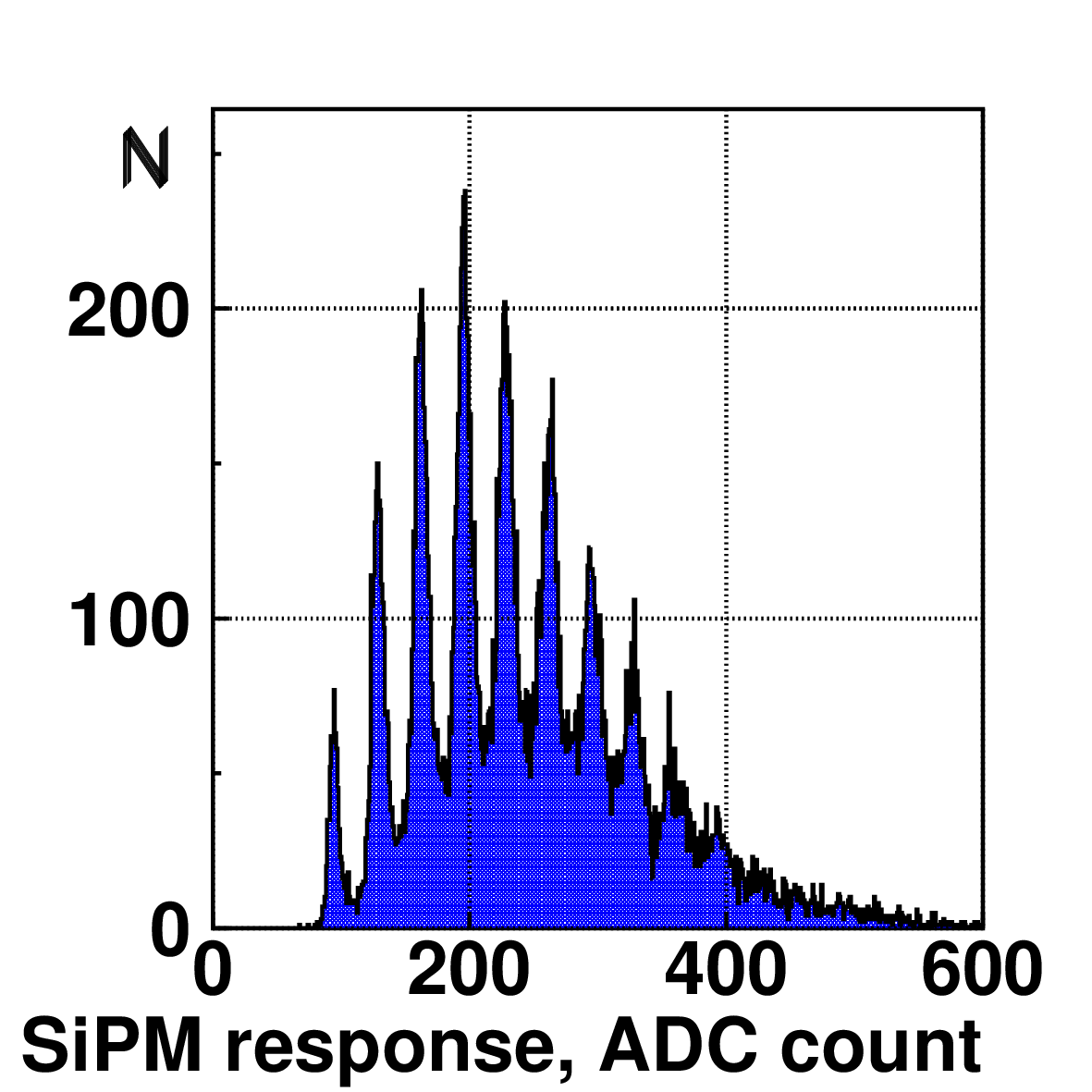}
\caption{A typical SiPM spectrum for low-intensity light, showing up
to nine individual peaks corresponding to different number of fired
pixels.}
\label{fig:sipm-pixel}
\end{figure}
This excellent resolution is extremely important for calorimetric
application since it allows a self calibration and monitoring of every
channel.  We record the response function of each SiPM over the entire
dynamic range (zero pixel to saturation). Figure~\ref{fig:saturation}
shows the number of fired pixels versus the light intensity in units
of MIPs for different SiPMs. 
\begin{figure}
\includegraphics[width=35mm]{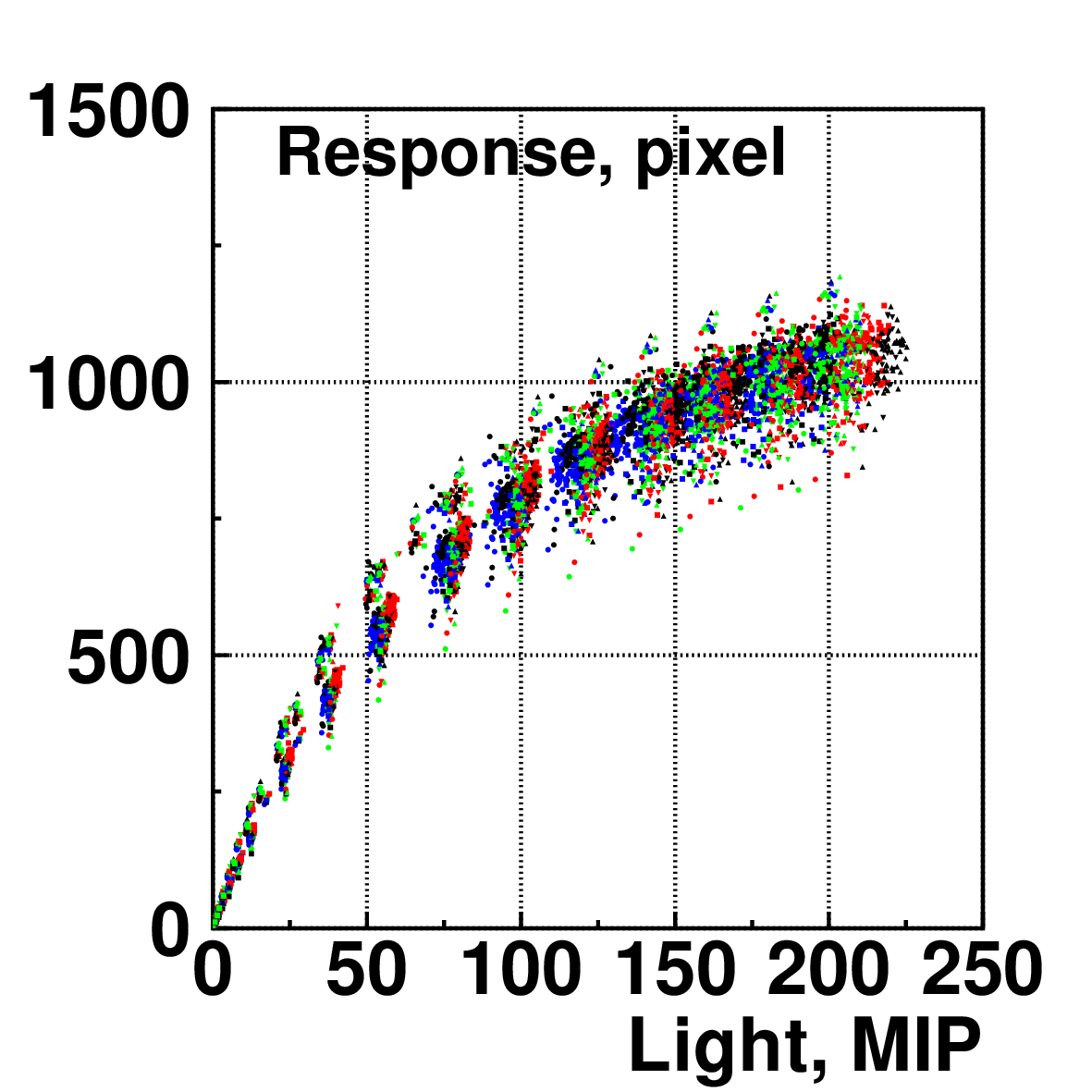}
\includegraphics[width=35mm]{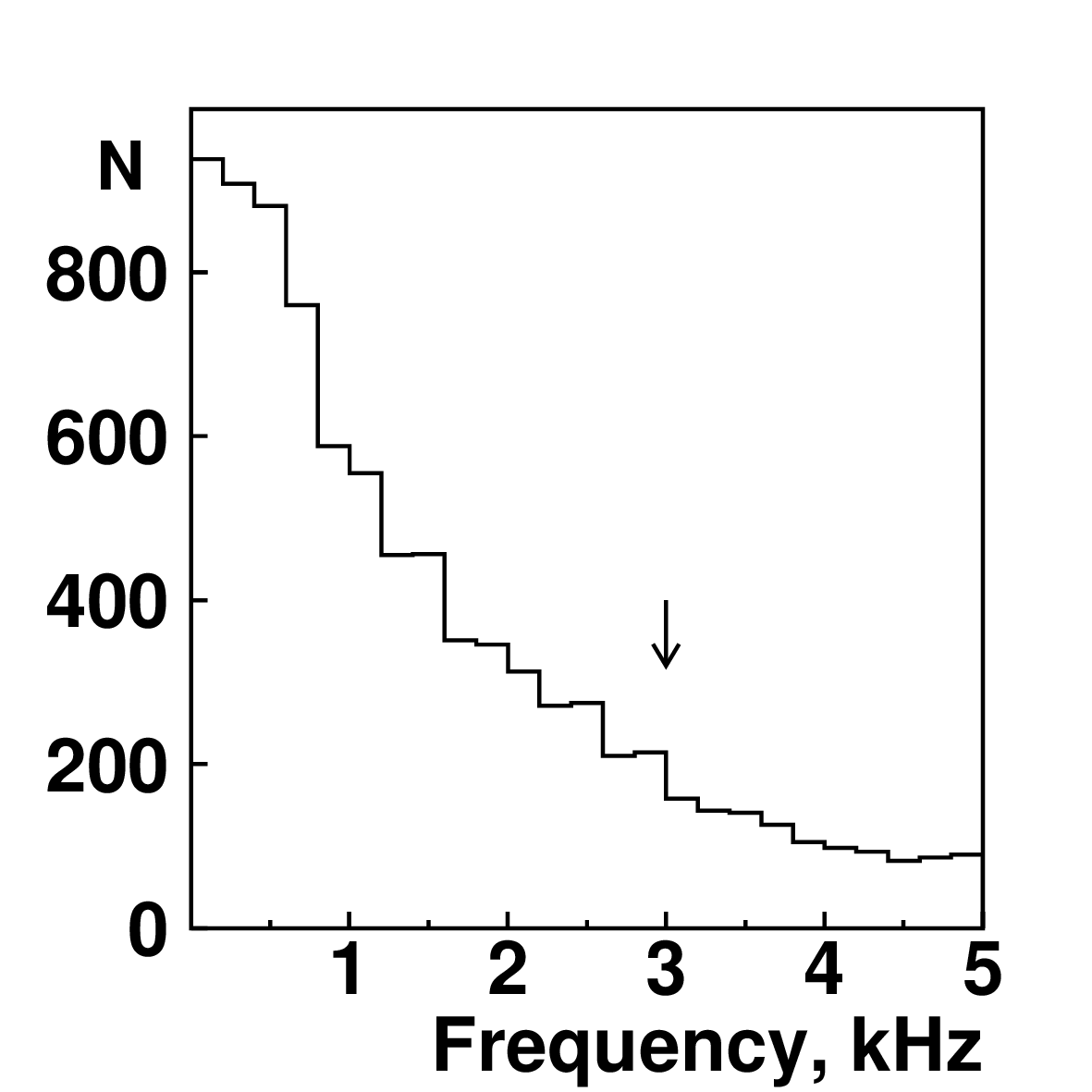}
\includegraphics[width=35mm]{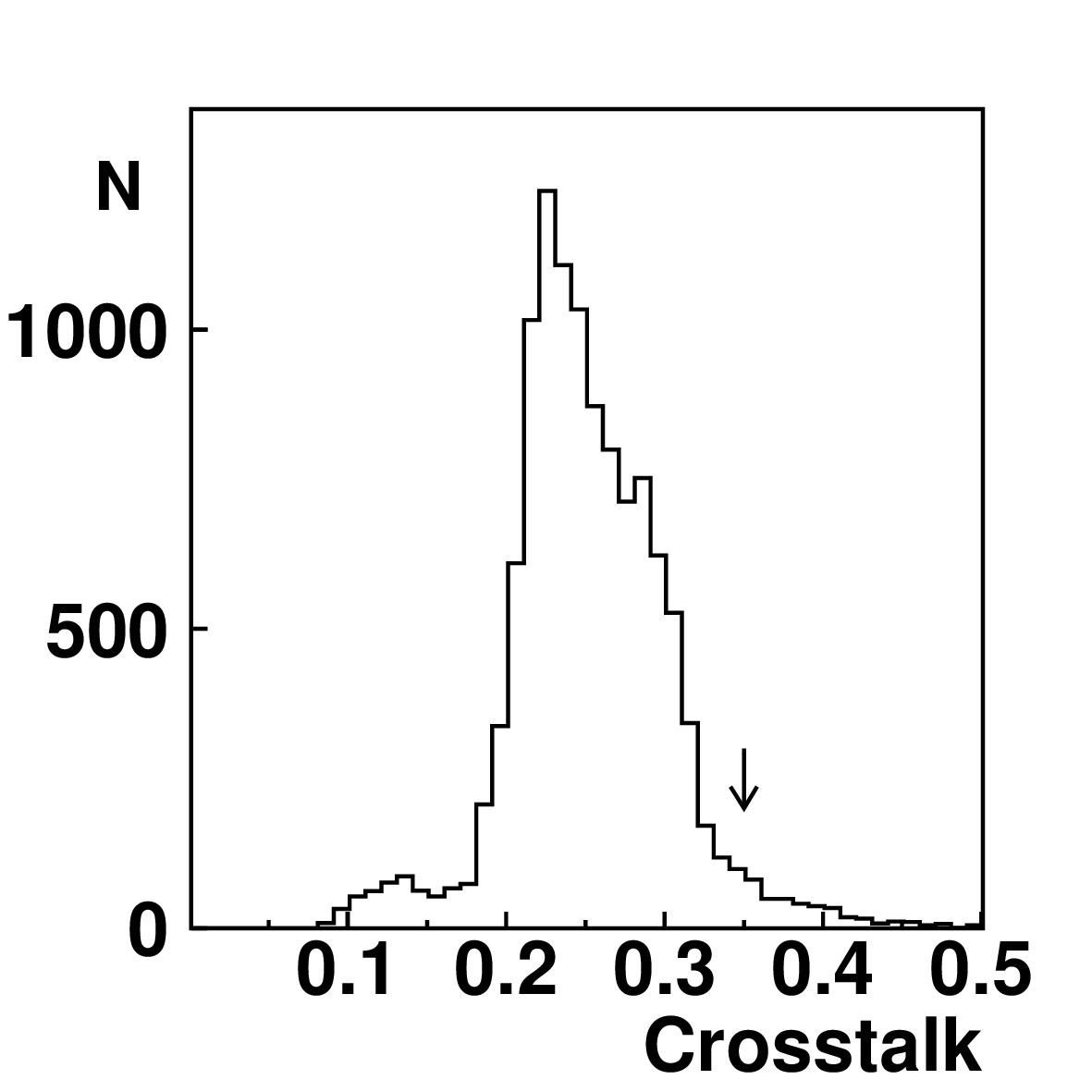}
\includegraphics[width=35mm]{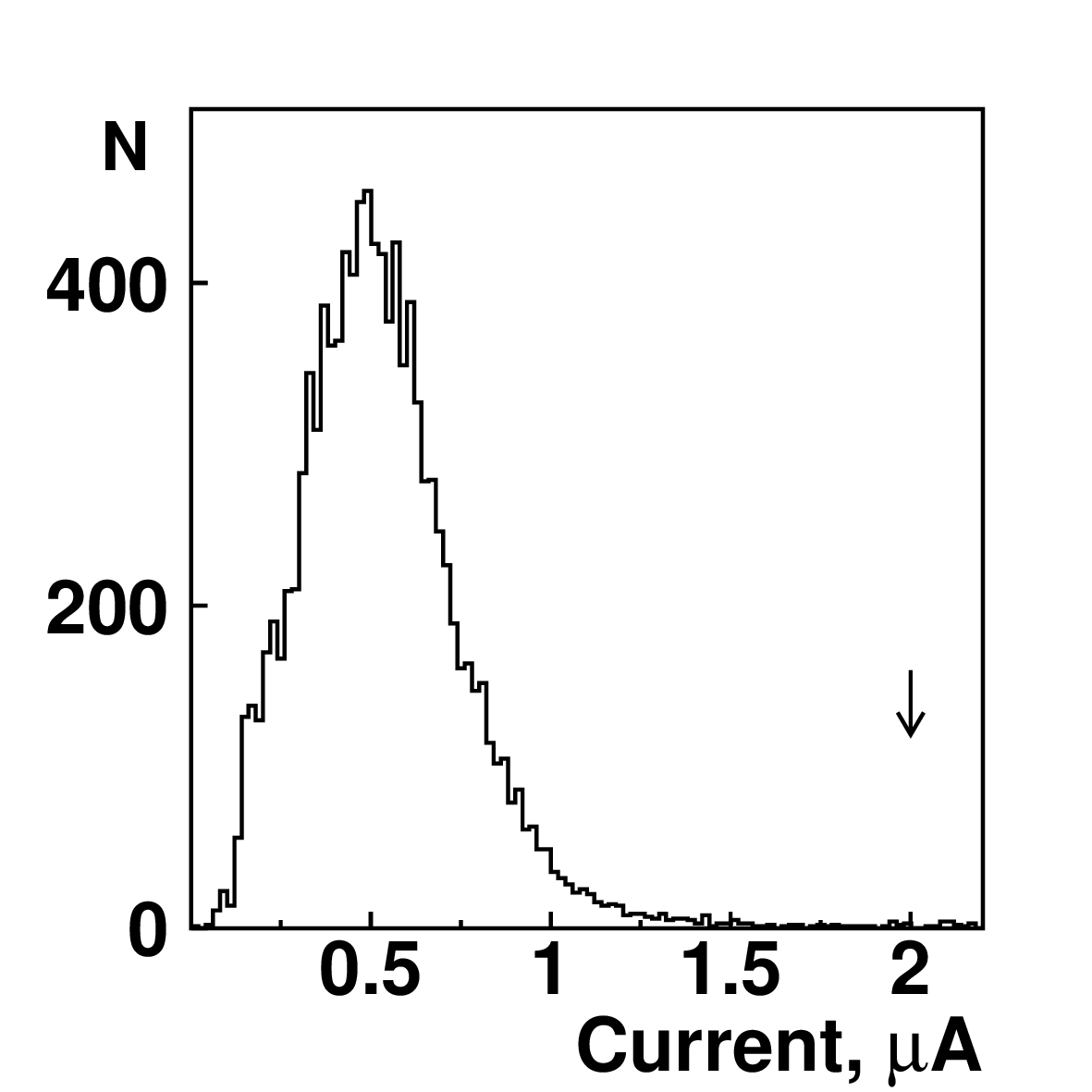}
\caption{The response function for SiPMs (upper left); the
distribution of SiPM noise at half a MIP threshold (upper right),
cross talk (lower left) and current (lower right).}
\label{fig:saturation}
\end{figure}
The shape of the response function of all SiPMs is similar and
individual curves are all within $Â±15\%$.  In addition, we measure
the noise rate at half a MIP threshold, the cross talk, and the SiPM
current. The corresponding distributions are also shown in figure
\ref{fig:saturation}. Arrows in figures show the selection cut-off.
Figure \ref{fig:spread} shows the SiPM parameter relative variation
for $0.1~V$ variation of the bias voltage.
\begin{figure}
\includegraphics[width=75mm]{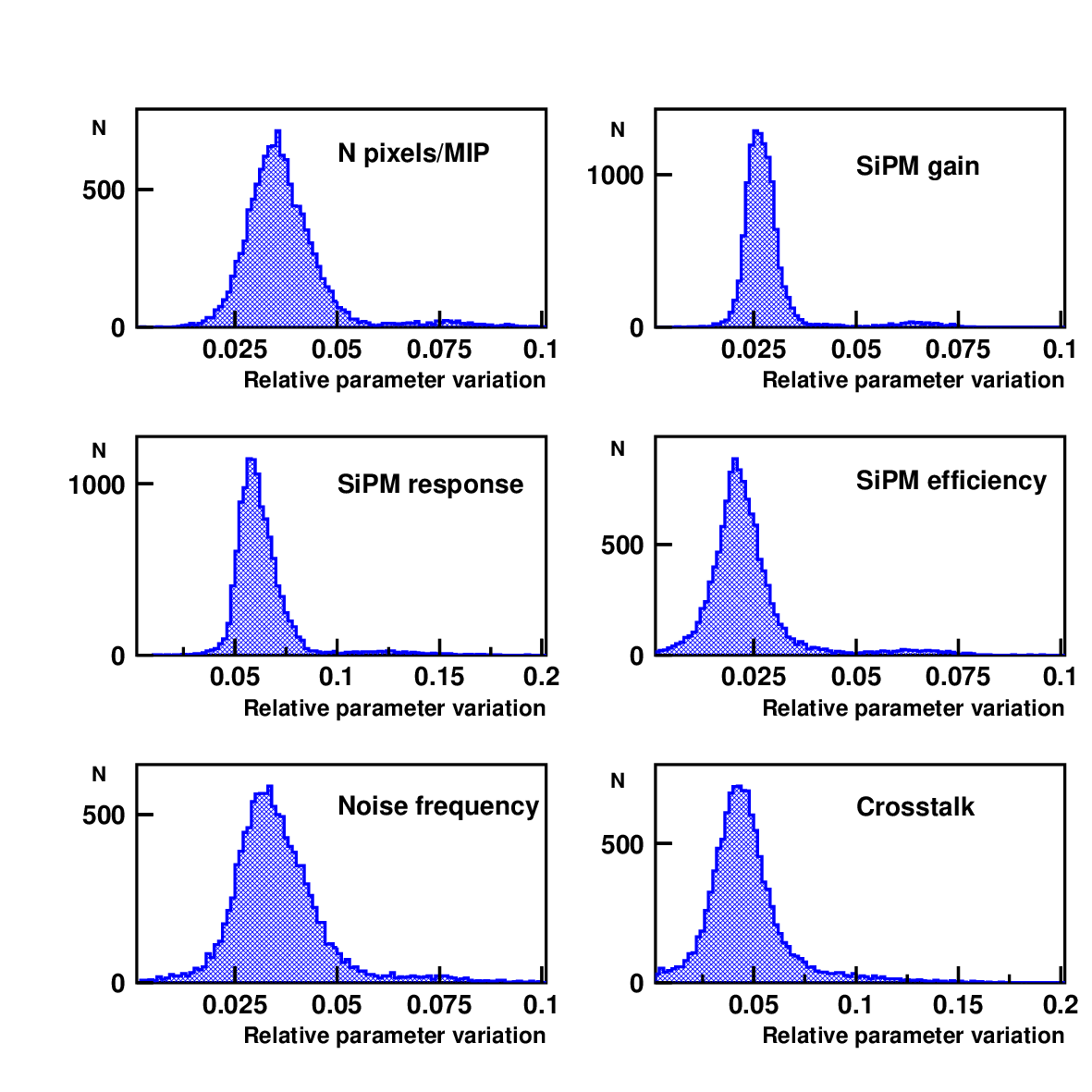}
\caption{The relative variation of SiPM parameters for 0.1~V change in
bias voltage.}
\label{fig:spread}
\end{figure}
The decrease of temperature by $2~^{\circ}$C leads to the decrease of
the breakdown voltage by $0.1~V$, which is equivalent to the increase
of the bias voltage by the same amount.  In addition, the decrease of
the temperature leads to the decrease of the SiPM noise.
 
The first radiation hardness tests of SiPMs have been performed using a
proton beam of the ITEP synchrotron. Figure~\ref{fig:protonrad} shows
the increase in SiPM current with the accumulated flux of 200~MeV
protons. 
\begin{figure}
\includegraphics[width=45mm]{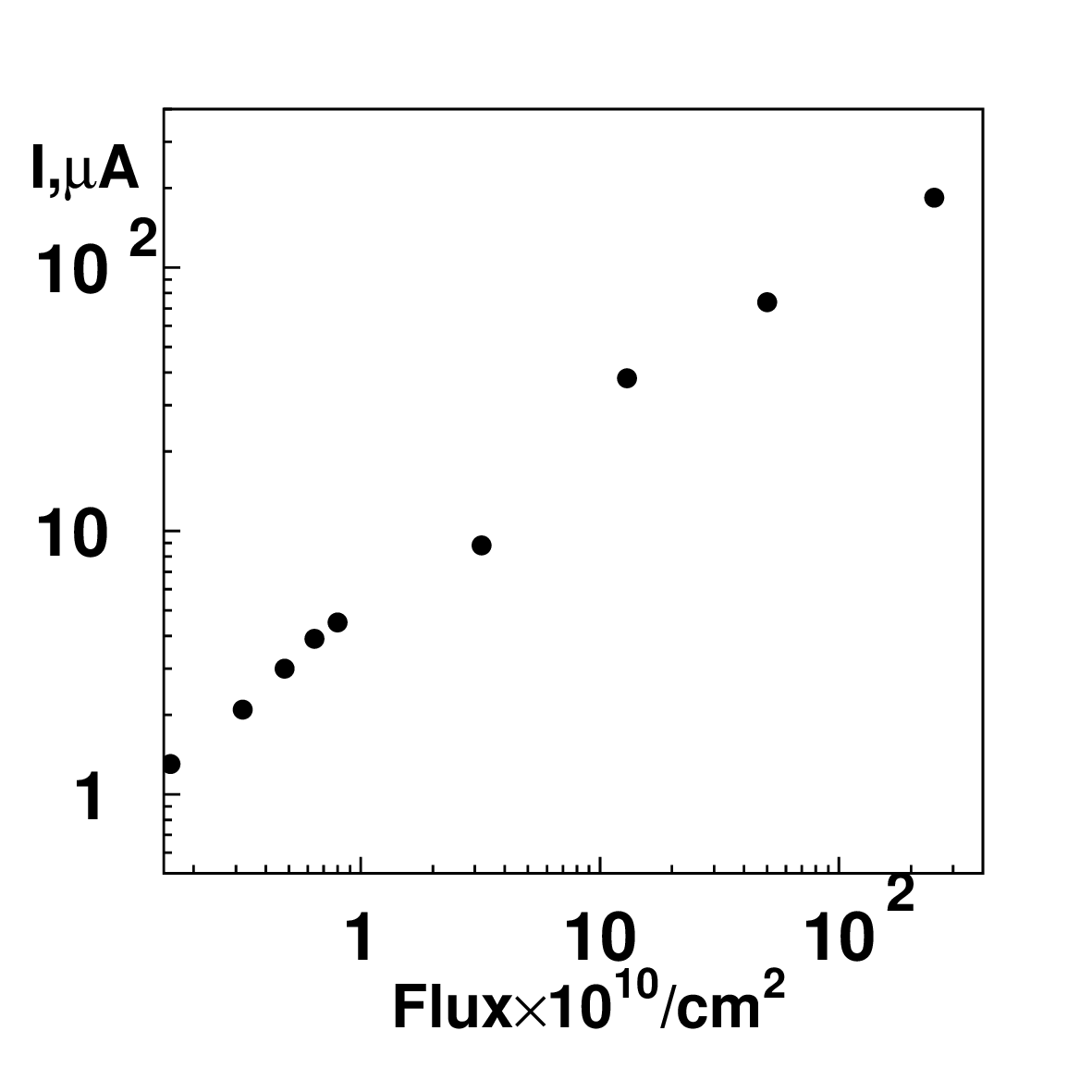}
\caption{The SiPM current as a function of flux of 200~MeV protons.}
\label{fig:protonrad}
\end{figure}
The current increase is compatible with that observed in
other Si detectors \cite{rad}:
\[
I=\alpha\cdot F\cdot S\cdot L\cdot G\cdot \epsilon\cdot (1+X)\,,
\]
where $F$ is the proton flux, $S$ is the SiPM area, $L$ is the
effective length in which noise charge carriers are produced,
$\epsilon$ is the probability for a noise carrier to produce the
Geiger discharge with the amplification factor $G$ and cross talk
$X$.  Using $\alpha=6\cdot10^{-17}~A/\cm$ \cite{rad} we get
$L\sim25\mu\m$.  SiPMs, however, are more sensitive to radiation
damage than other Si detectors because of the high amplification
($\sim10^6$) and a very low initial noise of about
0.1~photo-electrons. These two properties are important for a clear
separation of signals with different number of detected photons as
seen in figure~\ref{fig:sipm-pixel}.  This advantage which is
important for calibration is lost after an irradiation with about
$10^{10}$ protons/$\cm^2$, because individual pixel peaks cannot be
resolved any longer due to noise pile-up.  Nonetheless, SiPMs can be
still operated even after much larger radiation doses, but they have
an increased noise.  The radiation hardness of SiPMs is sufficient
for operation in a hadron calorimeter at the ILC. Only in the endcaps
close to the beam pipe, one can expect a neutron flux above $\rm
10^{10} /cm^2 /500~fb^{-1}$, which would lead to excess currents
above $\rm 5 \mu A$ and thus cause a smearing of individual pixel
peaks. We have assumed here a standard energy-dependent relative
radiation damage efficiency of neutrons and protons \cite{rad}.
Future tests will study the effect of long-term low-dose irradiation
on the aging.

During the production of the first two calorimeter planes we
encountered a problem of long not quenched discharges in many SiPMs
which appeared after some time. Problematic SiPMs have been studied
under a high gain microscope and the reason for the long discharges
was found. It was a short circuit between the polysilicon resistor and
the Al bus.  The distance between them is only $3~\mu\m$. This
distance will be increased in the next versions of SiPMs. For the
existing SiPMs we introduced the 48 hours test at elevated voltage
which removes the majority of problematic SiPMs.
     
\section{Test Beam Experience and Future Perspectives}

The calorimeter was assembled and commissioned at DESY.  The DESY
electron test beam was used to obtain an initial MIP calibration of
the calorimeter cells \cite{desy}.  The calorimeter was operated
practically without problems at the CERN test beam during 15 weeks
initially with 15 and then with 23 planes ($\sim 5000$ channels).  In
the planes 3-23 (which were produced after the observation of the long
discharge problem) 98\% of channels are good, 1\% are dead because of
the problems in SiPM-signal cable soldering, and 1\% are dead because
of the long discharges. We have not noticed any deterioration of the
calorimeter performance during the 15 weeks of operation.

The TCMT \cite{TCMT} is made of $100\times 5\times 0.5~\cm^3$ strips
with WLS fiber and SiPM readout.  All its 16 layers have been
installed and demonstrated a stable operation. This technique is a
good candidate for the ILC detector muon system
\cite{stripITEP}. Recent measurements of a
$100\times2.5\times0.5~\cm^3$ strip with WLS fiber and SiPM readout at
the ITEP synchrotron demonstrated a possibility to reach the time
resolution of better than 1~ns. This corresponds to the coordinate
resolution of about $15~\cm$ along the strip.
In the summer 2007 the complete HCAL and TCMT will be tested at CERN.

The ILC detector cost increases very fast with the increase of the
calorimeter thickness. Therefore it is desirable to have as thin
detector layers as possible.  We have studied at the proton beam of
the ITEP synchrotron $3~\mm$ thick $30\times30\mm^2$ tiles with both
WLS fiber, and direct coupling of MGPDs.  The direct coupling of MGPDs
to a tile simplifies considerably the production.
Figure~\ref{fig:lego} shows the tile response uniformity for WLS fiber
readout with SiPM and the direct readout with the $2.1\times
2.1~\mm^2$ blue-sensitive MRS APD (CPTA-149) from CPTA(Moscow)
\cite{CPTA}. 
\begin{figure}
\includegraphics[width=40mm]{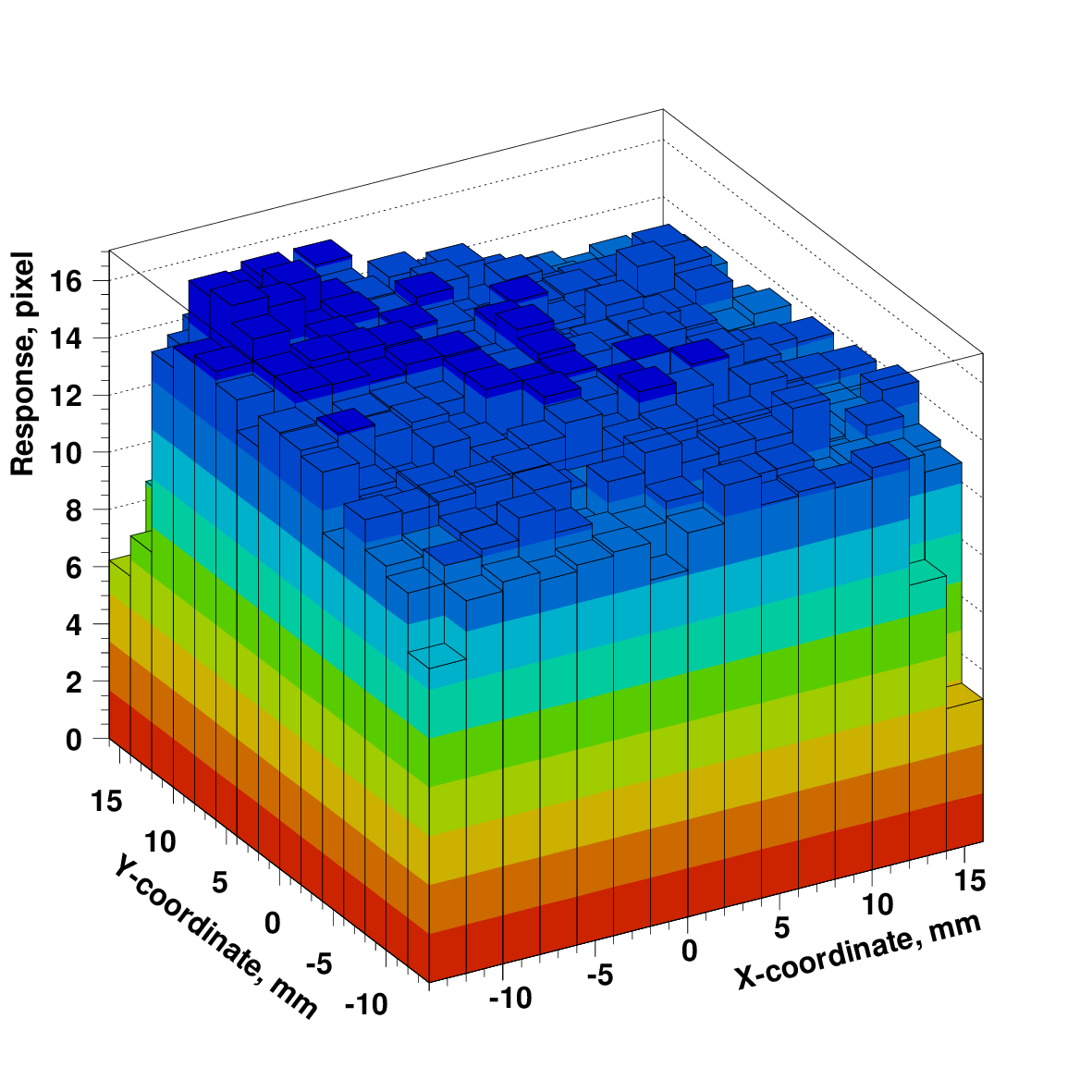}
\includegraphics[width=40mm]{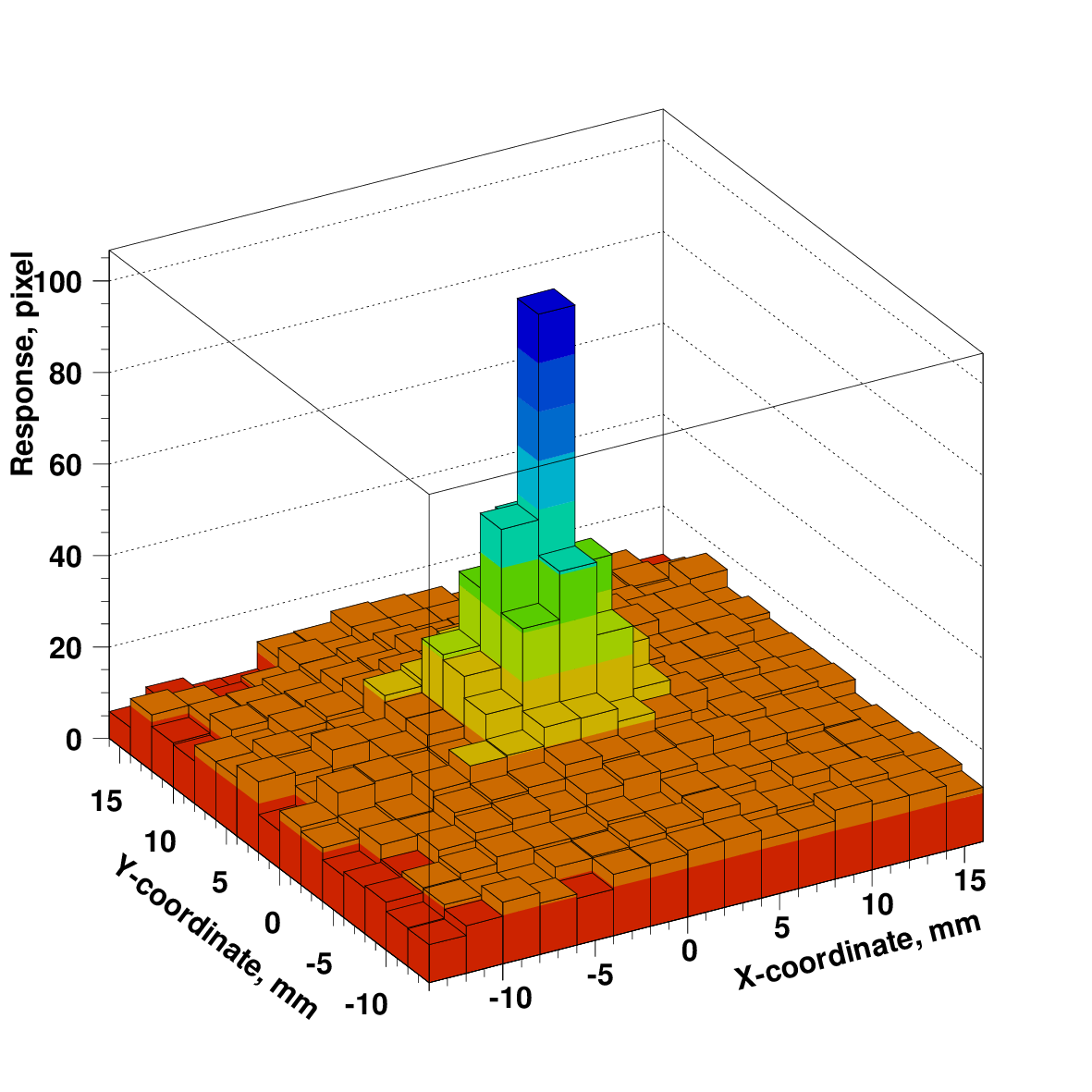}
\includegraphics[width=40mm]{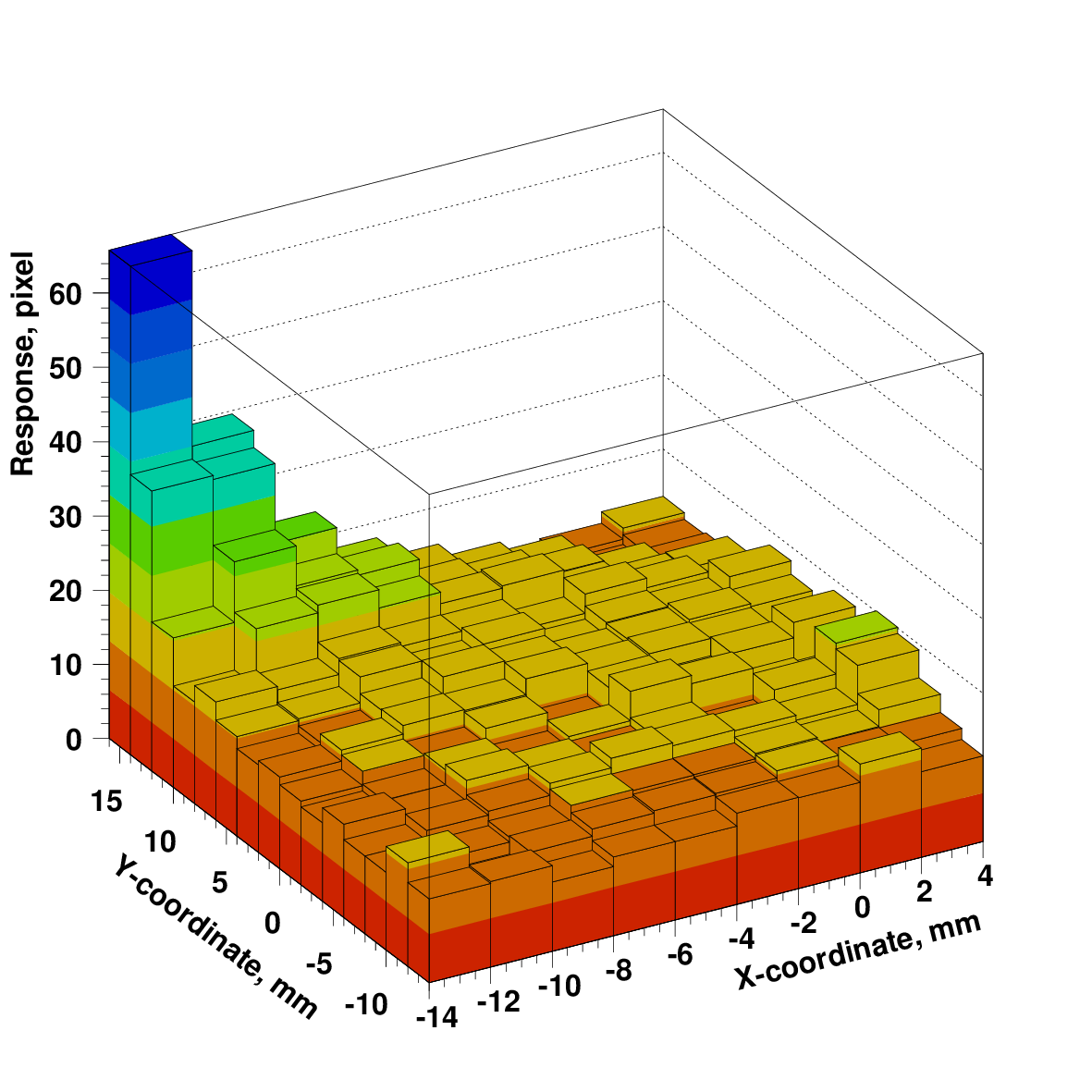}
\caption{$3\times30\times30\,\mm^3$ tile response to MIP; tile with a
diagonal fiber and SiPM readout (left), tiles with direct MRS~APD
readout (middle and right).}
\label{fig:lego}
\end{figure}
The Kuraray $1~\mm$ diameter Y11 fiber was glued inside $2~\mm$ deep
grove with an optical glue. The SiPM was also glued to the fiber.  The
MRS APD was attached with the optical glue at the center of the tile
or at the corner. The uniformity and the photo-electron yield are
sufficient for the tile with WLS fiber and SiPM readout.  There is a
large response non-uniformity in case of the direct MGPD
couplings. Probably it is possible to improve the uniformity by
reducing the reflectivity of the tile.  The uniformity requirements
for the hadron calorimeter are not high. The photo-electron yield in
the plateau region is sufficient. However the noise of the used MRS
APDs is too high to resolve individual photo-electron peaks. This
makes impossible the auto-calibration of the calorimeter. We consider
the auto-calibration to be extremely important for stability
monitoring and non-linearity correction.

Recently Hamamatsu has developed a MGPD called MPPC. The MPPC with
1600 channels has a very low noise and good efficiency in the blue
region.  Figure~\ref{fig:mgpd} shows our measurements of the
efficiency for green and blue light, gain, and cross-talk for
different MGPDs.  
\begin{figure}
\includegraphics[width=75mm]{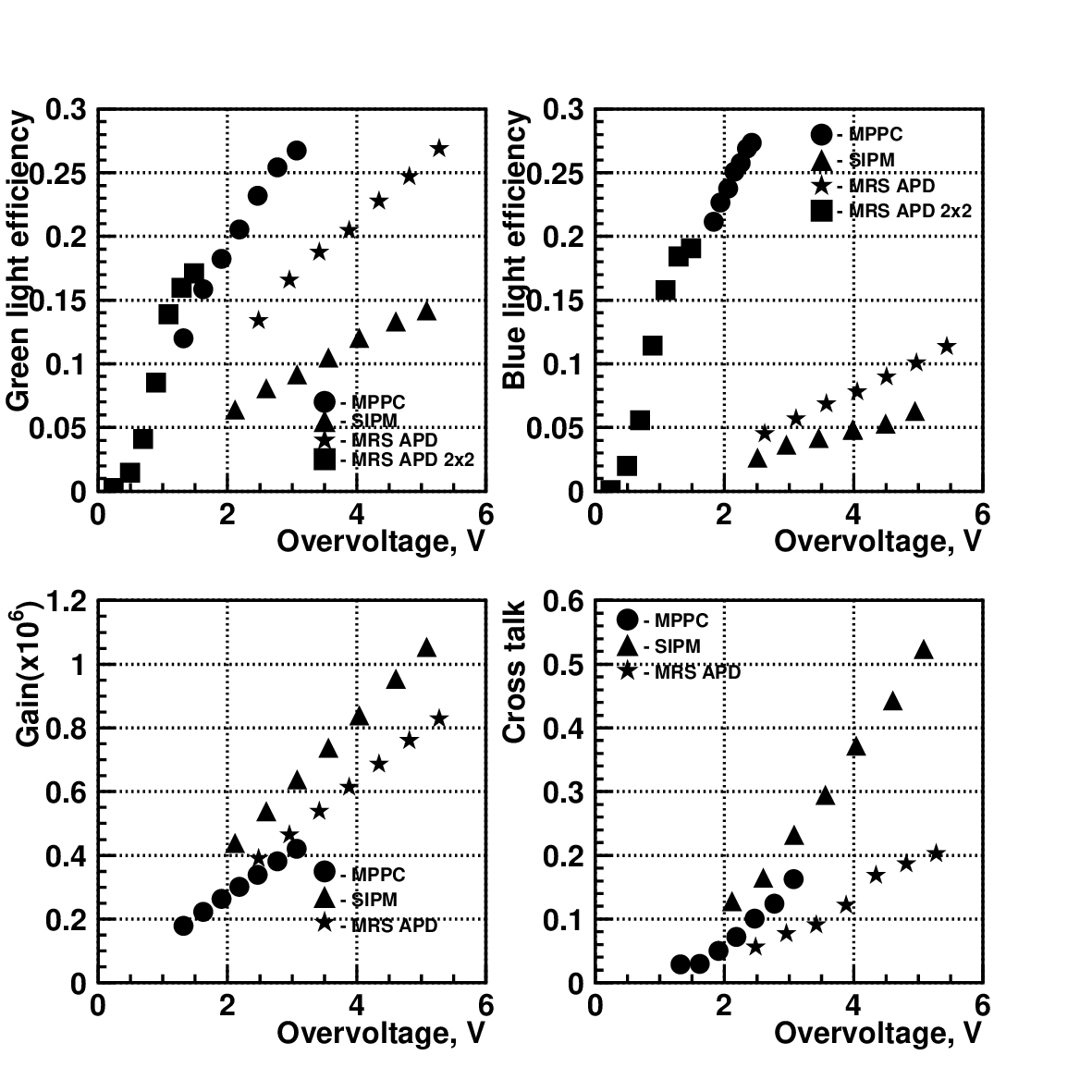}
\caption{Parameters of SiPM (triangles), MRS APD CPTA-143 (stars),
MRS APD CPTA-149 (squares), and MPPC (dots) as a function of
overvoltage.}
\label{fig:mgpd}
\end{figure}
The MPPC efficiency is measured with about $10\%$ accuracy. The SiPM
and MRS APD efficiencies are measured relative to the MPPC
efficiency. The green light is the light of a Y11 WLS fiber
illuminated with a UV LED. The blue light is the light of the
scintillator illuminated with the UV LED.  The green-sensitive
$1~\mm^2$ MRS APD (CPTA-143) and MPPC have a smaller cross talk and
larger efficiency for the green light than SiPMs.  The blue-sensitive
$4.4~\mm^2$ MRS APD (CPTA-149) has efficiency comparable to MPPC both
for the green and blue light. We could not measure the efficiency for
CPTA-149 at large overvoltages.  Individual pixels are not resolved in
this photo-detector.  Therefore the number of photo-electrons from the
LED is determined from the width of the pulse height distribution
assuming that it is determined by the Poison fluctuations in the
number of photo-electrons. This method stops to work at large
overvoltages because the noise becomes too large.  The MPPC has a
quite high efficiency for the blue light.  Unfortunately it is still
slightly insufficient for the direct readout of $3~\mm$ thick
tiles. The light yield measured at DESY and ITEP is about 7~pixels/MIP
for $5~\mm$ thick tiles and 5~pixels/MIP for $3~\mm$ thick
tiles. Since the MPPC noise is small the increase of the MPPC area by
a factor of two would be adequate for the direct readout of $3~\mm$
thick tiles. The MPPC has two small drawbacks. It has a relatively
small gain and a response curve which depends on the duration of the
signal.

\section{Conclusions}

The CALICE HCAL is the first large scale (8 thousand channels) application 
of the novel photo-detectors - SiPMs.
The unique SiPM properties and the elaborated test and selection procedure
 allowed to construct a reliable and simple in operation calorimeter.
The beam tests of HCAL will be very important for the demonstration 
of the feasibility of the PF method.
The developed technique looks adequate for a several hundred times larger
ILC hadron calorimeter but a lot of industrialization of this young 
technology is still required.
It is possible (and natural) to use the same technique for the ILC muon
system and may be even for the electromagnetic calorimeter.
The tile thickness can be reduced to $3~\mm$.
The feasibility of direct coupling of MGPD to a tile without WLS fiber
is still under investigation. New MGPDs are being developed by several firms.
The final choice of the photo-detector depends on the progress in the MGPD
properties and on the overall optimization.
The CALICE collaboration designs now a new realistic and scalable 
HCAL prototype.  
  
We would like to thank B.Dolgoshein, E.Garutti, S.Klemin, R.Mizuk,
E.Popova, V.Rusinov, F.Sefkow, and E.Tarkovsky for useful
discussions. This work has been partially supported by the Russian
Federal Atomic Energy Agency and ISTC grant 3090.

\end{document}